\documentclass{article}

\usepackage{frascatiphys-my}

\newcommand{\beq}{\begin{equation}}
\newcommand{\eeq}{\end{equation}}

\begin{document}

\title{Measuring quark polarizations at ATLAS and CMS}
\author{
Yevgeny Kats \\
{\em Department of Physics, Ben-Gurion University, Beer-Sheva~8410501, Israel}}
\maketitle
\baselineskip=10pt
\begin{abstract}
Being able to measure the polarization of quarks produced in various processes at the LHC would be of fundamental significance. Measuring the polarizations of quarks produced in new physics processes, once discovered, can provide crucial information about the new physics Lagrangian. In a series of recent papers, we have investigated how quark polarization measurements can be done in practice. The polarizations of heavy quarks ($b$ and $c$) are expected to be largely preserved in the lightest baryons they hadronize into, the $\Lambda_b$ and $\Lambda_c$, respectively. Furthermore, it is known experimentally that $s$-quark polarization is preserved as well, in $\Lambda$ baryons. We study how ATLAS and CMS can measure polarizations of $b$, $c$ and $s$ quarks using certain decays of these baryons. We propose to use the Standard Model $t\bar t$ and $Wc$ samples to calibrate these measurements. We estimate that the Run~2 dataset will suffice for measuring the quark polarizations in these Standard Model samples with precisions of order $10\%$. We also propose various additional measurements for the near and far future that would help characterize the polarization transfer from the quarks to the baryons.
\end{abstract}
\baselineskip=14pt

\section{Introduction}

ATLAS and CMS already measure the polarization of top quarks. In single top production, the tops are found to be highly polarized,\cite{Khachatryan:2015dzz,Aaboud:2017aqp} as expected from the parity-violating electroweak nature of the process, while tops from pair production are found to be unpolarized,\cite{Chatrchyan:2013wua,Aaboud:2016bit} as one indeed expects from production via QCD. Similarly, if top quarks from new physics processes are discovered, measuring their polarization will teach us about their production mechanism.

The question we have asked in three recent papers\cite{Galanti:2015pqa,Kats:2015cna,Kats:2015zth} was whether analogous polarization measurements could be done for quarks other than the top. The answer is, of course, not straightforward because these quarks are observed only as jets of hadrons. We have shown that it is nevertheless possible for ATLAS and CMS to measure the quark polarizations, as will be summarized in this note.

\section{Quark polarization retention in baryons}

\subsection{Bottom and charm quarks}

For heavy quarks, $m_q \gg \Lambda_{\rm QCD}$ (like the $b$ and $c$), the quark usually ends up in a very energetic heavy-flavored hadron\cite{Abbiendi:2002vt,Cacciari:2002re} that is easy to tell apart from the other hadrons in the jet. When this hadron is a baryon, an ${\cal O}(1)$ fraction of the quark polarization is expected to be retained in the hadron.\cite{Mannel:1991bs,Ball:1992fw,Falk:1993rf} Indeed, since the $b$-quark chromomagnetic moment $\mu_b \propto 1/m_b$, and $m_b \gg \Lambda_{\rm QCD}$, the $b$ spin should be approximately preserved during hadronization. If the light degrees of freedom of the baryon are in a spin-0 state, the $b$ spin is preserved also during the hadron's lifetime, $\tau$, despite the fact that $\tau \gg 1/\Lambda_{\rm QCD}$. Such a spin-1/2 baryon, with the valence light quarks being the $u$ and $d$ (which is the most common case), is called the $\Lambda_b$. The light degrees of freedom may also be in a spin-1 state. This gives rise to the $\Sigma_b$ and $\Sigma_b^\ast$ baryons, which have spins 1/2 and 3/2, respectively. In this case the $b$ spin oscillates during the baryon's lifetime. This is important to take into account because the $\Lambda_b$ sample has a significant contribution from $\Sigma_b^{(\ast)} \to \Lambda_b\pi$ (where the soft pion is difficult to identify, especially when it is neutral). The total fragmentation fraction $f(b \to {\rm baryons}) \approx 8\%$, and the baryons are almost entirely $\Lambda_b$, produced either directly or via $\Sigma_b^{(\ast)}$.

For the $c$ quark, one also has $m_c \gg \Lambda_{\rm QCD}$, although only as a rough approximation, and the same story holds. The analogs of the $\Sigma_b$ and $\Sigma_b^\ast$ baryons are often called the $\Sigma_c(2455)$ and $\Sigma_c(2520)$, respectively, but for the purpose of our discussion we will be referring to them as the $\Sigma_c$ and $\Sigma_c^\ast$. The fragmentation fraction to baryons in this case is $f(c \to {\rm baryons}) \approx 6\%$.

The dominant $b$-quark polarization loss effect in the $\Lambda_b$ baryons sample is expected to be due to the $\Sigma_b^{(\ast)} \to \Lambda_b\pi$ decays.\cite{Falk:1993rf} The quantity of interest, the polarization retention fraction
\beq
r \equiv \frac{{\cal P}(\Lambda_b)}{{\cal P}(b)} \,,
\eeq
can be expressed in terms of two parameters, $A$ and $w_1$,\cite{Falk:1993rf} which can be determined independently from other measurements, as described in the following. These parameters describe the probabilities for the heavy quark to hadronize in various ways. The parameter $A$ is the ratio of the $\Sigma_b$ and $\Sigma_b^\ast$ vs.\ direct $\Lambda_b$ production rates, which depends on the probability for the $b$ quark to hadronize with the spin-0 vs.\ spin-1 states of the light degrees of freedom. The parameter $w_1$ describes the probability for the spin-1 state of the light degrees of freedom to have polarization $\pm 1$ (as opposed to $0$) along the fragmentation axis. If one neglects the interference between the $\Sigma_b$ and $\Sigma_b^\ast$ states, the polarization loss effect is simply due to the fact that the process starts with the \emph{$b$ quark} having a definite spin, while when the hadron decays the wavefunction needs to collapse on a state of a definite \emph{hadron} spin (because the $\Sigma_b$ and $\Sigma_b^\ast$ decays can in principle be distinguished due to their different masses). Using the appropriate Clebsch-Gordan coefficients, one then obtains the polarization retention fractions
\beq
r_L \simeq \frac{1+(1+4w_1)A/9}{1+A}\;,\qquad\quad
r_T \simeq \frac{1+(5-2w_1)A/9}{1+A}
\eeq
for the cases in which the quark is polarized longitudinally and transversely, respectively, relative to the fragmentation axis. In practice, the $\Sigma_b$ and $\Sigma_b^\ast$ widths are not completely negligible relative to the mass splitting between them. With a more accurate calculation approach, which takes this into account,\cite{Galanti:2015pqa} one gets
\beq
r_L \approx \frac{1 + (0.23 + 0.38w_1)A}{1+A}\;, \qquad
r_T \approx \frac{1 + (0.62 - 0.19w_1)A}{1+A}
\label{rLrT-b}
\eeq
for the $b$-quark system. The analogous results for the $c$-quark system are
\beq
r_L \approx \frac{1 + (0.07 + 0.46w_1)A}{1+A}\;, \qquad
r_T \approx \frac{1 + (0.54 - 0.23w_1)A}{1+A}\;.
\label{rLrT-c}
\eeq
As we discuss in more detail in Ref.\cite{Galanti:2015pqa}, at the moment it is still difficult to extract precise and reliable values of $A$ and $w_1$ (for either the bottom or the charm system) from the various existing measurements or theoretical models. One can conclude from them, however, that $A \sim {\cal O}(1)$, while the value of $w_1$, for which the physically meaningful range is $0 \leq w_1 \leq 1$, is uncertain. This is sufficient for concluding from Eqs.~(\ref{rLrT-b}) and~(\ref{rLrT-c}) that the polarization retention fractions are ${\cal O}(1)$.

Notably, the ALEPH, DELPHI and OPAL experiments at LEP have attempted to measure the polarization transfer from quarks to baryons by analyzing the $\Lambda_b$ baryons in $Z \to b\bar b$ events, and have indeed found ${\cal O}(1)$ retention of the longitudinal polarization, although the statistical uncertainties were too large to extract a precise value of $r_L$.\cite{Buskulic:1995mf,Abreu:1999gf,Abbiendi:1998uz}

\subsection{Light quarks}

Moving to the strange quark, one cannot argue for polarization retention based on the heavy-quark picture. At the same time, one cannot argue for polarization loss either. And in fact $\Lambda$ polarization studies have been done in $Z$ decays at LEP, and found ${\cal O}(1)$ polarization retention for $\Lambda$ baryons that carry a significant fraction of the original quark's momentum.\cite{Buskulic:1996vb,ALEPH:1997an,Ackerstaff:1997nh} The dependence of the polarization transfer on the momentum fraction is described by the polarized (or spin-dependent) fragmentation functions,\cite{deFlorian:1997zj} which are universal functions, up to renormalization-group evolution, similar to the parton distribution functions.

Similarly to the strange quark, up and down quarks hadronizing to baryons are also expected to retain a fraction of their polarization. This will be difficult to measure, however, because the baryons most frequently produced in the $u$ and $d$ hadronization, the protons and neutrons, do not decay within the detector. One may still imagine using the $\Lambda$, for example, although that will require significantly more statistics.

\section{Opportunities at the LHC}

Nice Standard Model samples of highly-polarized quarks are available in $pp \to t\bar t$ events:\cite{Galanti:2015pqa,Kats:2015cna}
\begin{itemize}
\item The decays $t \to W^+b$ produce polarized $b$ quarks, and the subsequent decays $W^+ \to c\bar s, u \bar d$ produce polarized $c$, $s$, $u$, $d$ quarks.
\item It is easy to select a clean $t\bar t$ sample (e.g., in the lepton+jets channel).
\item Kinematic reconstruction of the event, along with charm tagging, enable obtaining separate samples of jets dominated by $b$, $c$, or $s$ jets.
\item Already by the end of Run~2, the statistics of polarized quarks from $t\bar t$ events will be as high as of those produced in the $Z$ decays at LEP.
\end{itemize}

Another potentially useful source of polarized charm quarks in the Standard Model is available in $pp \to Wc$ events (with $W \to \ell\nu$).\cite{Kats:2015zth} The statistics here are even higher than in $t\bar t$ (by an order of magnitude), but the backgrounds are higher too.

The simplest measurement to do is probably in $s$ jets (which can be selected as jets that accompany the charm-tagged jets in $t\bar t$ events), since one can use the low-background (thanks to the large displacement) and fully reconstructible decay
\beq
\Lambda \to p\,\pi^- \,.
\eeq
The $\Lambda$ polarization can be extracted from the angular distribution of the decay products, which is given by
\beq
\frac{1}{\Gamma}\frac{d\Gamma}{d\cos\theta} = \frac12\left(1 + \alpha{\cal P}(\Lambda)\cos\theta\right) \,,
\eeq
where $\alpha = 0.642 \pm 0.013$\cite{PDG} and $\theta$ is the angle (in the $\Lambda$ rest frame) between the proton momentum and the $\Lambda$ polarization. We estimate\cite{Kats:2015cna} that statistical precision of roughly $16\%$ is possible in ATLAS/CMS $t\bar t$ samples with $100$~fb$^{-1}$ of data.

The next simplest measurement is probably in $c$ jets, where one has the fully reconstructible, although not background-free, decay
\beq
\Lambda_c^+ \to p K^- \pi^+ \,,
\eeq
where the angular distribution is again sensitive to the polarization (see Ref.\cite{Kats:2015zth} for more details). We find that in $t\bar t$ samples, $100$~fb$^{-1}$ of data will allow achieving statistical precision of order $10\%$.\cite{Galanti:2015pqa} The reach of such a measurement in $W$+$c$ samples may be better or worse than in $t\bar t$, depending on the details.\cite{Kats:2015zth} It is interesting to note that the measurement we propose in the $W$+$c$ samples is quite similar to the existing measurements of the $W$+$c$ production cross section in ATLAS\cite{Aad:2014xca} and CMS\cite{Chatrchyan:2013uja}, which in particular rely on reconstructing the decay $D^+ \to K^-\pi^+\pi^+$, which is somewhat similar to our decay of interest. LHCb has analyzed the $W$+$c$ production process as well,\cite{Aaij:2015cha} and may also be able to measure the charm polarization.

In $b$ jets, the measurement is somewhat more complicated because the best decays to use seem to be the semileptonic decays (with a muon)
\beq
\Lambda_b \to X_c\, \mu^- \bar\nu \,,
\label{semileptonic}
\eeq
where $X_c$ stands for a charmed hadron (a typical example is the $\Lambda_c$) and any accompanying particles (e.g., pions).\cite{Galanti:2015pqa} The $\Lambda_b$ has several much cleaner decay channels, such as $\Lambda_b \to (J/\psi\to\mu^+\mu^-)\,(\Lambda\to p\pi^-)$, however their branching fractions are very small. Having a sizable branching fraction is an important consideration in our context because the most interesting potential application---new physics samples---will likely involve limited statistics. Additionally, the spin analyzing properties of the inclusive semileptonic decay in Eq.~(\ref{semileptonic}) are known accurately from theory.\cite{Manohar:1993qn,Czarnecki:1993gt,Czarnecki:1994pu}

Because of the invisible neutrino, and due to the difficulty in distinguishing between light neutral hadrons that are part of the $X_c$ and those coming from the primary vertex, the reconstruction of the $\Lambda_b$ here involves certain approximations. Additionally, there is a large intrinsic background due to the semileptonic decays of $B$ mesons. Even though the $B$-meson decay products are distributed isotropically in the meson rest frame, they contribute to the angular distribution measurements via statistical fluctuations. We have analyzed three different approaches for dealing with the $B$-meson background. In the first, \emph{exclusive} approach, we demand a fully-reconstructible $\Lambda_c$ decay candidate to be present in the jet. In the second, \emph{semi-inclusive} approach, we only require a reconstructed $\Lambda \to p\,\pi^-$ decay candidate to be present. In the third, \emph{inclusive} approach, no attempt is made to reduce the $B$-meson background, so the signal efficiency is maximal. We find that all the three approaches happen to offer similar levels of sensitivity, giving statistical precision of order $10\%$ for $100$~fb$^{-1}$ of $t\bar t$ data.

It is also interesting to consider the dominant production mechanism of $b$ quarks at the LHC, namely the inclusive QCD production, $pp \to b\bar b+X$. Despite the enormous cross section of this process, it is not the most promising avenue for polarization measurements because these quarks are produced unpolarized at the leading order. However, a small transverse polarization is predicted at the next-to-leading order.\cite{Dharmaratna:1996xd,Bernreuther:1995cx} This, in principle, provides an opportunity to measure the transverse polarization transfer, namely the parameter $r_T$.\footnote{One should be careful with the interpretation though, since transverse polarization in baryons is not protected from being generated by soft QCD effects independent of the polarization of the original quark.\cite{Mulders:1995dh,Anselmino:2000vs}} It is not a simple measurement because the transverse polarization is suppressed at high momenta, ${\cal P}(b) \sim \alpha_s m_b/p_b$, and has a strong dependence on the kinematics of the parton-level process. There already exist analyses in LHCb\cite{Aaij:2013oxa} and CMS\cite{Sirunyan:2018bfd} (and a related analysis in ATLAS\cite{Aad:2014iba}) that have attempted to detect the $\Lambda_b$ transverse polarization (using the very rare but clean decay channel $\Lambda_b \to J/\psi\,\Lambda$), although they are subobtimal due to their inclusiveness over the kinematics. More detailed studies will be possible with higher statistics.

Finally, we note that the parameters $A$ and $w_1$ that enter the expressions for $r_L$ and $r_T$ in Eqs.~(\ref{rLrT-b}) and~(\ref{rLrT-c}) can be measured independently, as has been pointed out already in Ref.\cite{Falk:1993rf}, even in samples of unpolarized $b$ or $c$ quarks, such as the inclusive QCD samples available to the LHC experiments. The measurements can be done in any experiment that can reconstruct the $\Sigma_b^{(\ast)} \to \Lambda_b\pi$ or $\Sigma_c^{(\ast)} \to \Lambda_c\pi$ decays, and involve measuring the production rates of these baryons (for determining $A$) and the angular distribution of the pion (for determining $w_1$). In fact, it might have been possible to do such measurements even at the Tevatron.\cite{CDF:2011ac} Measurements of the $\Sigma_c^{(\ast)}$ can be done also at $B$ factories, as demonstrated by the recently released analysis from Belle,\cite{Niiyama:2017wpp} from which one may infer $A \approx 0.5$ for the charm system.

\section{Conclusions}

Our work\cite{Galanti:2015pqa,Kats:2015cna,Kats:2015zth} motivates a number of experimental analyses:

\begin{enumerate}

\item In $t\bar t$ samples in ATLAS and CMS:

\begin{itemize}
\item Longitudinal $\Lambda_b$ polarization measurement in $b$ jets from top decays. This will allow determining $r_L$ for the bottom.
\item Longitudinal $\Lambda_c$ polarization measurement in $c$ jets from $W$ decays. This will allow determining $r_L$ for the charm.
\item Longitudinal $\Lambda$ polarization measurement in $s$ jets from $W$ decays. This will provide information about the longitudinally-polarized $s \to \Lambda$ fragmentation function.
\item In the far future: longitudinal $\Lambda$ polarization measurement in $u$ and $d$ jets from $W$ decays. This will provide information about the longitudinally-polarized $u \to \Lambda$ and $d \to \Lambda$ fragmentation functions.
\end{itemize}

\item In $W(\to\ell\nu)+c$ samples in ATLAS, CMS, and perhaps LHCb:

\begin{itemize}
\item Longitudinal $\Lambda_c$ polarization measurement in the $c$ jets. This will allow determining $r_L$ for the charm.
\item LHCb in particular may attempt separating out the $\Sigma_c^{(\ast)} \to \Lambda_c\pi$ contributions.
\end{itemize}

\item In QCD production of hard quarks in ATLAS, CMS, and LHCb:

\begin{itemize}
\item Transverse $\Lambda_b$ (maybe also $\Lambda_c$) polarization measurement, properly binned in the event kinematics. This may allow determining $r_T$ for the bottom (charm).
\end{itemize}

\item In any (even unpolarized) samples of hard quarks in LHCb, ATLAS and CMS:

\begin{itemize}
\item Measurement of the $\Sigma_b^{(\ast)}$ production yields (relative to direct $\Lambda_b$ production), and the pion angular distribution in the $\Sigma_b^\ast$ decays. This will allow determining the parameters $A$ and $w_1$, respectively, for the bottom quark.
\item Measurement of the $\Sigma_c^{(\ast)}$ production yields (relative to direct $\Lambda_c$ production), and the pion angular distribution in the $\Sigma_c^\ast$ decays. This will allow determining the parameters $A$ and $w_1$, respectively, for the charm quark.
\end{itemize}

\item In new-physics samples, once discovered by ATLAS and/or CMS:

\begin{itemize}
\item 
Measurements of the final-state quark polarizations. The results will provide important information about the structure of the new-physics interactions.

We note however, that given that no new physics has been discovered so far, and considering the price one needs to pay in fragmentation and branching fractions to measure quark polarizations, statistics will likely be a serious limitation for such measurements.
\end{itemize}

\item In $t\bar t$ and $Wc$ production, in the long term, in ATLAS, CMS, and LHCb:

\begin{itemize}
\item Measurements of the full polarized fragmentation functions for the various quark flavors. It will be possible to confront the results with models based on the heavy-quark effective theory for the $b$ and $c$ quarks, and more phenomenological models of QCD for the light quarks. Additionally, knowing the full fragmentation functions will allow computing the scale dependence (due to the renormalization group evolution)\cite{Stratmann:1996hn} of the polarization retention fractions.
\end{itemize}

\end{enumerate}

\section{Acknowledgements}

I would like to thank the organizers of the LFC17 workshop at ECT* for their invitation to present these ideas and for a stimulating workshop. I am also grateful to Mario Galanti, Andrea Giammanco, Yuval Grossman, Emmanuel Stamou and Jure Zupan for the collaboration on the work that started this research program.

\end{document}